  \documentclass{cjour}
\usepackage{amsmath}
\usepackage[T1]{fontenc}
\usepackage{epsfig}
\usepackage{times}

\newcommand{\sil}{\:\lower0.6ex\hbox{$\stackrel{\textstyle
<}{\sim}$}\:}

\begin{document}



\authorrunninghead{W. Hillebrandt et al.}
\titlerunninghead{Thermonuclear explosions}






\title{Thermonuclear Explosions of Chandrasekhar-Mass White Dwarfs}

\subtitle{Thermonuclear Explosions}

\author{Wolfgang Hillebrandt and Martin Reinecke}
\affil{Max-Planck-Institut f\"ur Astrophysik, D-85740 Garching, Germany}
\and
\author{Jens C. Niemeyer}
\affil{LASR, Enrico Fermi Institute, Univ. of Chicago, Chicago IL 60637, USA}



\dedication{Dedicated to Joe Monaghan on the occasion of his 60th birthday}

\abstract{
We present a new way of modeling turbulent thermonuclear
deflagration fronts in Chandrasekhar-mass white dwarfs,  
consisting of carbon and oxygen, undergoing a type Ia
supernova explosion. Our approach is a front capturing/tracking
hybrid scheme, based on a level set method,
which treats the front as a mathematical discontinuity and
allows for full coupling between the front geometry and the flow
field. First results of the method applied to the problem of 
type Ia supernovae are discussed. It will be shown that even in
2-D and even with a physically motivated sub-grid model
numerically ``converged'' results are difficult to obtain.
}


\begin{article}



\section{Introduction}
Type Ia supernovae, i.e. stellar explosions which do not have hydrogen
in their spectra, but intermediate-mass elements, such as silicon,
calcium, cobalt, and iron, have recently received considerable
attention because it appears that they can be used as ``standard
candles'' to measure cosmic distances out to billions of light years
away from us. Moreover, observations of type Ia supernovae seem to
indicate that we are living in a universe that started to accelerate
its expansion when it was about half its present age. These
conclusions rest primarily on phenomenological models which, however,
lack proper theoretical understanding, mainly because the explosion
process, initiated by thermonuclear fusion of carbon and oxygen into
heavier elements are difficult to simulate even on supercomputers, 
for reasons we shall discuss in this article.
 
The most popular progenitor model for the average type Ia supernovae 
is a massive white dwarf, consisting of carbon and oxygen,
which approaches the Chandrasekhar mass, M$_{\text{Ch}}$
$~\simeq~\text{1.39}$~M$_{\odot}$,
by a yet unknown mechanism, presumably accretion from a companion star, and
is disrupted by a thermonuclear explosion (see, e.g., \cite{W90}
for a review). Arguments in favour of this hypothesis include the ability of 
these models to fit the observed spectra and light curves.

However, not only is the evolution of massive white dwarfs 
to explosion very uncertain, leaving room for some diversity 
in the allowed set of initial conditions (such as the temperature 
profile at ignition), but also the physics of thermonuclear burning
in degenerate matter is complex and not well understood. The generally
accepted scenario is that explosive carbon burning is ignited 
either at the center of the star or off-center in a couple of ignition
spots, depending on the details of the previous evolution. 
After ignition, the flame is thought to
propagate through the star as a sub-sonic deflagration wave which may or
may not change into a detonation at low densities (around 10$^7$g/cm$^3$).
Numerical models with parameterized velocity of the burning front
have been very successful, the prototype being the W7 model of
\cite{NTY84}. However, these models do
not solve the problem because attempts to determine the effective
flame velocity from direct numerical simulations failed and gave
velocities far too low for successful explosions \cite{L93,
K95, AL94a}. This has led to some speculations about ways
to change the deflagration into a supersonic detonation \cite{K91a,
K91b}.

Numerical simulations of any kind of 
turbulent combustion have always been a challenge,
mainly because of the large range of length scales involved. In type
Ia supernovae, in particular, the length scales of relevant
physical processes range from 10$^{-3}$cm for the Kolmogorov-scale
to several 10${^7}$cm for typical convective motions.
In the currently favored scenario the explosion starts as a deflagration
near the center of the star. Rayleigh-Taylor unstable blobs of hot burnt
material are thought to rise and to lead to shear-induced turbulence
at their interface with the unburnt gas. This turbulence increases the
effective surface area of the flamelets and, thereby, the rate of fuel
consumption; the hope is that finally a fast deflagration 
might result, in agreement with phenomenological models of type
Ia explosions \cite{NTY84}.

Despite considerable progress in the field of modeling turbulent
combustion for astrophysical flows (see, e.g., \cite{NH95a}),
the correct numerical representation of the thermonuclear deflagration front
is still a weakness of the simulations. Methods used up to now are
based on for the reactive-diffusive flame model \cite{K93a},
which artificially stretches the burning region over several grid zones 
to ensure an isotropic flame propagation speed. 
However, the soft transition from fuel to ashes stabilizes
the front against hydrodynamical instabilities on small length scales,
which in turn results in an underestimation of the flame surface area and
--~consequently~-- of the total energy generation rate. Moreover,
because nuclear fusion rates depend on temperature nearly
exponentially, one cannot use the zone-averaged values of the
temperature obtained this way to calculate the reaction kinetics.

The front tracking method described here cures some of these 
weaknesses. It is based on the so-called
\emph{level set technique} which was originally  
introduced by Osher and Sethian \cite{OS88}. They used
the zero level set of a $n$-dimensional scalar function to represent
$(n-1)$-dimensional front geometries. Equations for the time evolution of
such a level set which is passively advected by a flow field are given in
\cite{SSO94}. The method has been extended to allow the tracking of
fronts propagating normal to themselves, e.g. deflagrations and detonations
\cite{SMK97}. In contrast to the
artificial broadening of the flame in the reaction-diffusion-approach, this
algorithm is able to treat the front as an exact hydrodynamical discontinuity.

In the following sections we present the 
main ideas and governing equations of this approach. 
We describe only the most simple
implementation of the flame model and leave extensions including
the reconstruction of the thermo-dynamical properties
of mixed cells to subsequent papers. The main emphasis will be to
demonstrate that the method can indeed be applied to the supernova problem.
In addition, we will show that 
even if one attempts to model the physics of thermonuclear burning on 
unresolved scales well by physically motivated LES 
one still has to perform calculations with very high
spatial resolution in order to be at least near to a converged solution.
 
The numerical results presented here were obtained by performing 2D
simulations, but there is no problem extending them to
3D. Such simulations are under way.
 
\section{General properties of exploding white dwarfs}

To a very good approximation, the exploding white dwarf material can
be  described as a fully ionized plasma with varying degrees of
electron degeneracy, satisfying the fluid approximation.  The
governing equations are the hydrodynamical equations for mass,
species, momentum, and energy transport including gravitational
acceleration, viscosity, heat and mass diffusion \cite{LL95}, and
nuclear energy generation \cite{A96}. They must  be supplemented by
an equation of state for an ideal gas of nuclei, an arbitrarily
relativistic and degenerate electron gas, radiation, and
electron-positron pair production and annihilation \cite{CG68}. The
gravitational potential is calculated with the help of the Poisson
equation. In numerical simulations that fully resolve the relevant
length scales  for dissipation, diffusion, and nuclear burning it is
possible to obtain the energy generation rate from a nuclear reaction
network (for a recent overview, see \cite{T99}) and the diffusion
coefficients from an evaluation 
of the kinetic transport mechanisms \cite{NP84}. If, on the other
hand, these scales are unresolved -- as  is usually the case in
simulations on scales of the stellar radius -- subgrid-scale models
are required to compute (or parameterize) the effective large-scale
transport coefficients and burning rates, which are more or less
unrelated to the respective microphysical quantities \cite{K95,NH95a}.

Initial conditions can be obtained from hydrostatic spherically
symmetric models of the accreting white dwarf or -- for Chandrasekhar
mass progenitors -- from the Chandrasekhar equation for a fully
degenerate, zero temperature white dwarf \cite{KW89}.  Given the
initial conditions and symmetries specifying the boundary conditions,
the dynamics of the explosion can in principle be determined by
numerically integrating the equations of motion. \cite{M98} gives a
detailed account of some current numerical techniques used for modeling
supernovae.

\subsection{Nuclear burning in degenerate C+O matter}
\label{flame}

Owing to the strong temperature dependence of the nuclear reaction
rates, $\dot S \sim T^{12}$ at $T \approx 10^{10}$ K \cite{HK94},
nuclear burning during the 
explosion is  confined to microscopically thin layers that propagate
either conductively as subsonic deflagrations (``flames'') or by shock
compression as supersonic detonations
\cite{CF48,LL95}. Both modes are hydrodynamically unstable
to spatial perturbations as can be shown by linear perturbation
analysis.  In the nonlinear regime, the burning fronts are either
stabilized by forming a cellular structure or become fully turbulent
-- either way, the total burning rate increases as a result of flame
surface growth \cite{LE61,W85,ZBLe85}. Neither flames nor detonations
can be resolved in explosion simulations on stellar scales and
therefore have to be represented by numerical models.

When the fuel exceeds a critical temperature $T_{\rm c}$ where burning
proceeds nearly instantaneously compared with the fluid motions (see
\cite{TW92} for a suitable definition of $T_{\rm c}$), a thin
reaction zone forms at the interface between burned and unburned
material. It propagates into the surrounding fuel by one of two
mechanisms allowed by the Rankine-Hugoniot jump conditions: a
deflagration (``flame'') or a detonation.

If the overpressure created by the heat of the burning products is
sufficiently high, a hydrodynamical shock wave forms that ignites the
fuel by compressional heating. Such a self-sustaining combustion front that
propagates by shock-heating is called a detonation. 
If, on the other hand, the initial overpressure is too weak, the
temperature gradient at the fuel-ashes interface steepens until an
equilibrium between heat diffusion (carried out predominantly by
electron-ion collisions) and energy generation is reached.  The
resulting combustion front consists of a diffusion zone that heats up
the fuel to $T_{\rm c}$, followed by a thin reaction layer where the
fuel is consumed and energy is generated. It is called a deflagration
or simply a flame and moves subsonically with respect to the unburned
material \cite{LL95}.  Flames, unlike detonations, may therefore be
strongly affected by turbulent velocity fluctuations of the fuel. Only
if the unburned material is at rest, a unique laminar flame speed
$S_{\rm l}$ can be found which depends on the detailed interaction of
burning and diffusion within the flame region, e.g.
\cite{ZBLe85}. Following \cite{LL95}, it can be
estimated by assuming that in order for burning and diffusion to be in
equilibrium, the respective time scales, $\tau_{\rm b} \sim
\epsilon/\dot w$ and $\tau_{\rm d} \sim \delta^2/\kappa$, where
$\delta$ is the flame thickness and $\kappa$ is the thermal
diffusivity, must be similar: $\tau_{\rm b} \sim \tau_{\rm
d}$. Defining $S_{\rm l} = \delta/\tau_{\rm b}$, one finds $S_{\rm l}
\sim (\kappa \dot w/\epsilon)^{1/2}$, where $\dot w$ should be
evaluated at $T \approx T_{\rm c}$ \cite{TW92}.  This is only a crude
estimate due to the strong $T$-dependence of $\dot w$. Numerical
solutions of the full equations of hydrodynamics including nuclear
energy generation and heat diffusion are needed to obtain more
accurate values for $S_{\rm l}$ as a function of $\rho$ and fuel
composition.  Laminar thermonuclear carbon and oxygen flames at high
to intermediate densities were investigated by
\cite{BCM80,IIC82,WW86b}, and, using a variety of different
techniques and nuclear networks, by \cite{TW92}. For the purpose of
SN Ia explosion modeling, one needs to know the laminar flame speed
$S_{\rm l} \approx 10^7 \dots 10^4$ cm s$^{-1}$ for $\rho \approx 10^9
\dots 10^7$ g cm$^{-3}$, the flame thickness $\delta = 10^{-4} \dots
1$ cm (defined here as the width of the thermal pre-heating layer
ahead of the much thinner reaction front), and the density contrast
between burned and unburned material $\mu = \Delta \rho/\rho = 0.2
\dots 0.5$ (all values quoted here assume a composition of $X_{\rm C}
= X_{\rm O} = 0.5$ \cite{TW92}). The thermal expansion  parameter
$\mu$ reflects the partial lifting of electron degeneracy in the
burning products, and is much lower than the typical value found in
chemical, ideal gas systems \cite{W85}.

Observed on scales much larger than $\delta$, the internal
reaction-diffusion  structure can be neglected and the flame can be
approximated as a density jump that propagates locally with the normal
speed $S_{\rm l}$. This ``thin flame'' approximation allows a linear
stability analysis of the front with respect to spatial
perturbations. The result shows that thin flames are linearly unstable
on all wavelengths. It was discovered first by Landau (1944) \cite{L44} and
Darrieus (1938) \cite{D38} and is hence called the ``Landau-Darrieus'' (LD)
instability.  Subject to the LD instability, perturbations grow until
a web of cellular  structures forms and stabilizes the front at
finite perturbation amplitudes \cite{Z66}. The LD instability
therefore does not, in   general, lead to the production of
turbulence. In the context of SN Ia models, the nonlinear LD
instability was studied by \cite{BS96}, using a statistical approach
based on the Frankel equation, and by \cite{NH95b} employing 2D
hydrodynamics and a one-step burning rate. Both groups concluded that
the cellular stabilization mechanism precludes a strong acceleration
of the burning front as a result of the LD instability. However,
\cite{BS96} mention the possible breakdown of stabilization at low
stellar densities (i.e., high $\mu$) which is also indicated by the
lowest density run of \cite{NH95b}.

\subsection{Hydrodynamic instabilities and turbulence}
 
The best studied and probably most important hydrodynamical effect for
modeling SN Ia explosions is the Rayleigh-Taylor (RT) instability
resulting from the buoyancy of hot, burned fluid with
respect to the dense, unburned material \cite{MA82,MA86,L93,K94,K95,NH95a},
and after more than five decades of
experimental and numerical work, the basic phenomenology of nonlinear
RT mixing is fairly well understood \cite{F51,L55,S84,R84,Y84}:
Subject to the RT instability, small surface perturbations grow until
they form bubbles (or ``mushrooms'') that begin to float upward while
spikes of dense fluid fall down. In the nonlinear regime, bubbles of
various sizes interact and create a foamy RT mixing layer whose
vertical extent $h_{\rm RT}$ grows with time $t$ according to a
self-similar growth law, $h_{\rm RT} = \alpha g (\mu/2) t^2$, where
$\alpha$  is a dimensionless constant ($\alpha \approx 0.05$) and $g$
is the background gravitational acceleration. 

Secondary instabilities related to  the velocity shear along the
bubble surfaces \cite{NH97} quickly lead to the production of  turbulent velocity
fluctuations that cascade from the size of the largest bubbles
($\approx 10^7$ cm) down to the microscopic Kolmogorov scale, $l_{\rm
k} \approx 10^{-4}$ cm where they are dissipated
\cite{NH95a,K95}. Since no computer is capable of resolving this
range of scales, one has to resort to statistical or scaling
approximations of those length scales that are not properly
resolved. The most prominent scaling relation in turbulence 
research is Kolmogorov's law for the cascade of velocity fluctuations,
stating that in the case of isotropy and statistical stationarity, the
mean velocity $v$ of turbulent eddies with size $l$ scales as $v \sim
l^{1/3}$ \cite{K41}. 
Given the velocity of large eddies, e.g. from
computer simulations, one can use this relation to extrapolate the
eddy velocity distribution down to smaller scales under the assumption
of isotropic, fully developed turbulence \cite{NH95a}.
Knowledge of the eddy velocity as a function of length scale is
important to classify the  burning regime of the turbulent combustion
front \cite{NW97,NK97,KOW97}. The ratio of the laminar flame speed
and the turbulent velocity on the scale of the flame thickness, $K =
S_{\rm l}/v(\delta)$, plays an important role: if  $K \gg 1$, the
laminar flame structure is nearly unaffected by turbulent
fluctuations. Turbulence does, however, wrinkle and deform the flame
on  scales $l$ where $S_{\rm l} \ll v(l)$, i.e. above the  {\em Gibson
scale} $l_{\rm g}$ defined by $S_{\rm l} = v(l_{\rm g})$
\cite{P88}. These  wrinkles increase the flame surface area and
therefore the total energy generation rate of the turbulent front
\cite{D40}. In other words, the turbulent  flame speed, $S_{\rm t}$,
defined as the mean overall propagation velocity of the turbulent
flame front, becomes larger than the laminar speed $S_{\rm l}$. If the
turbulence is sufficiently strong,  $v(L) \gg  S_{\rm l}$, the
turbulent flame speed becomes independent of the laminar speed, and
therefore of the microphysics of burning and diffusion, and scales
only with the velocity of the largest turbulent eddy \cite{D40,C94}:
\begin{equation}
\label{st}
S_{\rm t} \sim v(L)\,\,.
\end{equation}
Because of the unperturbed laminar flame properties on very small
scales, and the wrinkling of the flame on large scales, the burning
regime where $K \gg 1$ is called the corrugated flamelet regime
\cite{P90,C94}. 

As the density of the white dwarf material declines and the laminar
flamelets become slower and thicker, it is plausible that at some
point turbulence significantly alters the thermal flame structure
\cite{KOW97,NW97}. This marks the end of the flamelet regime and
the beginning of the distributed burning, or distributed reaction
zone, regime, e.g. \cite{P90}. So far, modeling the distributed burning
regime in exploding white dwarfs has not been attempted explicitely
since neither nuclear burning and diffusion nor turbulent mixing can
be properly described by simplified prescriptions. Phenomenologically,
the laminar flame structure is believed to be disrupted by turbulence
and to form a distribution of reaction zones with various lengths and
thicknesses. In order to find the critical density for the transition
between both regimes, we need to formulate a specific criterion for
flamelet breakdown.  A criterion for the transition between both
regimes is discussed in \cite{NW97,NK97} and \cite{KOW97}:
\begin{equation}
l_{\rm cutoff} \le \delta\,\,.
\end{equation} 
Inserting the results of \cite{TW92} for $S_{\rm l}$ and $\delta$ as
functions of density, and using a typical turbulence velocity
$v(10^6\mbox{cm}) \sim 10^7$ cm s$^{-1}$, the transition from flamelet
to distributed burning can be shown to occur at a density of
$\rho_{\rm dis} \approx 10^7$ g cm$^{-3}$ \cite{NK97}.

\section{A numerical scheme for turbulent combustion}
 
Our aim is now to convert some of the ideas presented in the previous 
Sections into a numerical scheme. The basic ingredients will be
a finite-volume method to solve the fluid-dynamics equation, a front-tracking
algorithm which allows us to propagate the thermonuclear flame (assumed
to be in the flamelet regime), and a model to determine
the turbulent velocity fluctuations on unresolved sub-grid scales.
 
\subsection{The level set method}
 
The central aspect of our front tracking method is the association
of the front geometry (a time-dependent set of points $\Gamma$)
with an isoline of a so-called level set function $G$:
\begin{equation}
  \Gamma:=\lbrace\vec r\ |\ G(\vec{r}) = 0\rbrace
\end{equation}
Since $G$ is not completely determined by this equation, we can additionally
postulate that $G$ be negative in the unburnt and positive in the burnt
regions, and that $G$ be a ``smooth'' function, which is convenient from
a numerical point of view. This smoothness can be achieved, for example,
by the additional constraint that $|\vec{\nabla}G| = 1$
in the whole computational domain, with the exception of possible extrema
and kinks of
$G$. The ensemble of these conditions produces a $G$ which is a signed distance
function, i.e. the absolute value of $G$ at any point equals the minimal
front distance. The normal vector to the front is defined to
point towards the unburnt material.

The task is now to find an equation for the temporal evolution of $G$
such that the zero level set of $G$ behaves exactly
as the flame. Such an expression can be obtained by the consideration
that the total velocity of the front consists of two independent contributions:
it is advected by the fluid motions at a speed $\vec v$ and
it propagates normal to itself with a burning speed $s$.

Since for deflagration waves a velocity jump usually
occurs between the pre-front and post-front states, we must explicitly
specify which state $\vec v$ and $s$ refer to; traditionally, the values
for the unburnt state are chosen. Therefore, one obtains for the total
front motion
\begin{equation}
\label{Df}
  \vec D_f = \vec v_u + s_u \vec n.
\end{equation}
The total temporal derivative of $G$ at a point $P$ attached to the front
must vanish, since $G$ is, by definition, always 0 at the front:
\begin{equation}
  \frac{\text{d}G_P}{\text{d}t} = \frac{\partial G}{\partial t}
  + \vec \nabla G \cdot \dot {\vec x}_P = \frac{\partial G}{\partial t}
  + \vec D_f\cdot \vec\nabla G = 0
\end{equation}
This leads to the desired differential equation describing the time
evolution of $G$:
\begin{equation}
  \frac{\partial G}{\partial t} = - \vec D_f\cdot \vec\nabla G.
  \label{levprop}
\end{equation}

This equation, however, cannot be applied on the whole computational
domain, main\-ly because  using this equation everywhere will
in most cases destroy $G$'s distance function property.
Therefore additional measures must be taken in the regions away from the front
to ensure a ``well-behaved'' $|\vec\nabla G|$ \cite{RHNe99}.

The situation is further complicated by the fact that the quantities
$\vec v_u$ and $s_u$ which are needed to determine $\vec D_f$ are not
readily available in the cells cut by the front. In a finite volume context,
these cells contain a mixture of pre- and post-front states instead.
Nevertheless one can assume that the conserved quantities (mass, momentum and
total energy) of the mixed state satisfy the following conditions:
\begin{alignat}{2}
  \overline{\rho}   &= \alpha \rho_u &&+ (1-\alpha) \rho_b \label{cons1}\\
  \overline{\rho \vec v} &= \alpha \rho_u \vec v_u
    &&+ (1-\alpha) \rho_b \vec v_b \label{cons2} \\
  \overline{\rho e} &= \alpha \rho_u e_u &&+ (1-\alpha) \rho_b e_b \label{cons3}
\end{alignat}
Here $\alpha$ denotes the volume fraction of the cell occupied by the unburnt
state.
In order to reconstruct the states before and behind the flame, a nonlinear
system consisting of the equations above, the Rankine-Hugoniot jump conditions
and a burning rate law must be solved. 

  \begin{figure}
  \centerline{\epsfig{file=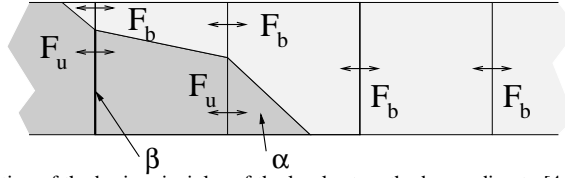,width=0.6\textwidth}}
  \caption{
   Illustration of the basic principles of the level set method according
   to \cite{SMK97}: The piecewise linear front cuts the mixed
   cells into burnt and unburnt parts. $\alpha$ is the unburnt volume fraction
   of a cell, $\beta$ is the unburnt area fraction of a cell interface. The
   fluxes $\vec F_u$ and $\vec F_b$ are calculated from the reconstructed
   states.}
  \label{flspl}
  \end{figure}

Having obtained the reconstructed pre- and post-front states in the mixed
cells, it is not only possible to determine $\vec D_f$, but also to separately
calculate the fluxes of burnt and unburnt material over the cell interfaces.
Consequently, the total flux over an interface can be expressed as a linear
combination of burnt and unburnt fluxes weighted by the unburnt interface area
fraction $\beta$ (see Fig. \ref{flspl}):
\begin{equation}
  \label{spliteq}
  \vec{\bar{F}} = \beta \vec F_u + (1-\beta)\vec F_b.
\end{equation}

\subsection{Implementation}
For our calculations, the front tracking algorithm was implemented as
an additional module for the hydrodynamics code PROMETHEUS
\cite{FMA89}. Here we describe a simple implementation of
most of the ideas discussed in the previous section which we will call
``passive implementation''. It assumes that the $G$-function
is advected by the fluid motions and by burning and is only used to
determine the source terms for the reactive Euler equations. 
It must be noted that there exists no \emph{real} discontinuity between
fuel and ashes in this case; the transition is smeared out over
about three grid cells by the hydro-dynamical scheme, and the level set
only indicates where the thin flame front \emph{should} be. However,
the numerical flame is still considerably thinner than in the
reaction-diffusion approach.

A complete implementation would contain in-cell-reconstruction and
flux-splitting as proposed by \cite{SMK97} 
and outlined above. Therefore it
should exactly describe the coupling between the flame and the hydrodynamic
flow. This generalized version of the code has so-far only
been applied to hydrogen combustion in air and, therefore, will be
discussed elsewhere.

\subsubsection{$G$-Transport}
Since the front motion consists of two distinct contributions,
it is appropriate to use an operator splitting approach for the
time evolution of $G$. The advection term due to the fluid velocity
$\vec v_F$ reads in conservation form
\begin{equation}
\int_V\frac{\partial (\rho G)}{\partial t}d^3 r
+\oint_{\partial V} -\vec v_F \rho G d\vec f = 0
\end{equation}
\cite{MOS92}. This equation is identical to the advection equation
of a passive scalar, like the concentration of an inert chemical species.
Consequently, this contribution to the front propagation can be calculated
by PROMETHEUS directly.
The additional flame propagation due to burning is calculated at the end
of each time step and a re-initiali\-zation of $G$ is done in order to
keep it a signed distance function (see \cite{RHNe99}).

\subsubsection{Source terms}
After the update of the level set function in each time step, the change
of chemical composition and total energy due to burning is calculated in the
cells cut by the front. In order to obtain these values, the volume fraction
$\alpha$ occupied by the unburnt material is determined in those cells by the
following approach: from the value $G_{ij}$ and the two steepest gradients
of $G$ towards the front in $x$- and $y$-direction a first-order approximation
$\tilde G$ of the level set function is calculated; then the area fraction
of cell $ij$ where $\tilde G < 0$ can be found easily.
Based on these results, the new concentrations of fuel, ashes and energy are
obtained:
\begin{align}
X'_{\text{Ashes}} &= \text{max}(1-\alpha, X_{\text{Ashes}}) \label{newash} \\
X'_{\text{Fuel}} &= 1-X'_{\text{Ashes}} \\
e'_{\text{tot}} &= e_{\text{tot}} + q (X'_{\text{Ashes}}-X_{\text{Ashes}})
\end{align}
In principle this means that all fuel found behind the front is converted
to ashes and the appropriate amount of energy is released. The maximum operator
in eq. (\ref{newash}) ensures that no ``reverse burning'' (i.e. conversion from
ashes to fuel) takes place in the cases
where the average ash concentration is higher than the burnt volume fraction;
such a situation can occur in a few rare cases because of unavoidable
discretization errors of the numerical scheme.

\subsection{Turbulent nuclear burning}
The system of equations described so-far can be solved provided the
normal velocity of the burning front is known everywhere and at all
times. In our computations it is determined according to a flame-brush
model of Niemeyer and Hillebrandt \cite{NH95a},
which we will briefly outline for convenience.

As was mentioned before, nuclear burning in degenerate dense matter
is believed to propagate on microscopic scales as a conductive
flame, wrinkled and stretched by local turbulence, but with essentially
the laminar velocity. Due to the very high Reynolds numbers macroscopic 
flows are highly turbulent and they interact with the flame,
in principle down to the Kolmogorov scale. This means
that all kinds of hydrodynamic instabilities feed energy into a
turbulent cascade, including the buoyancy-driven Rayleigh-Taylor
instability and the shear-driven Kel\-vin-Helm\-holtz instability.
Consequently, the picture that emerges is more that of a ``flame
brush'' spread over the entire turbulent regime rather than a wrinkled
flame surface. For such a flame brush, the relevant minimum length
scale is the so-called Gibson scale, defined as the lower bound for
the curvature radius of flame wrinkles caused by turbulent stress.
Thus, if the thermal diffusion scale is much smaller than the Gibson
scale (which is the case for the physical conditions of interest here)
small segments of the flame surface are unaffected by large scale
turbulence and behave as unperturbed laminar flames (``flamelets''). 
On the other hand side, since the Gibson scale is, at high densities,
several orders of magnitude smaller than the integral scale set by
the Rayleigh-Taylor eddies and many orders of magnitude larger than
the thermal diffusion scale, both transport and burning times are
determined by the eddy turnover times, and the effective velocity of
the burning front is independent of the laminar burning velocity.

A numerical realization of this general concept is presented in
\cite{NH95a}. The basic assumption
was that wherever one finds turbulence this turbulence is fully
developed and homogeneous, i.e. the turbulent velocity fluctuations on
a length scale $l$ are given by the Kolmogorov law 
$v(l) = v(L) (l/L)^{1/3}$, 
where $L$ is the integral scale, assumed to be equal to the
Rayleigh-Taylor scale. Following the ideas outlined above, one can
also assume that the thickness  of the turbulent flame brush on the scale
$l$ is of the order of $l$ itself. With these two assumptions and the
definition of the Gibson scale one finds
 for $l_{\text{gibs}} \sil l \sil L \simeq \lambda _{\text{RT}}$
\begin{equation} 
\label{vtur}
v(l) \simeq u_t(l) \simeq u_t(l_{\text{gibs}}) \biggl(\frac{l}{l_{\text{gibs}}} 
\biggr)^{1/3}  
\end{equation}
and $d_t(l) \simeq l$, 
where $v(l_{\text{gibs}}) = u_{\text{lam}}$ defines $l_{\text{gibs}}$,
$u_{\text{lam}}$ is the
laminar burning speed and $u_t(l)$ is the
turbulent flame velocity on the scale $l$.

In a second step this model of turbulent combustion is coupled to our finite
volume hydro scheme. 
Since in every finite volume scheme scales smaller than the grid size 
cannot be resolved, we express $l_{\text{gibs}}$ in terms of the grid size
$\Delta$, the (unresolved) turbulent kinetic energy $q$, and the
laminar burning velocity:
\begin{equation}
\label{gibs}
l_{\text{gibs}} = \Delta \biggl( \frac{u_l^2}{2q} \biggr) ^{3/2}.
\end{equation}
Here $q$ is determined from a sub-grid model
\cite{C93,NH95a} and, finally, the effective
turbulent velocity of
the flame brush on scale $\Delta$ is given by 
\begin{equation}
\label{vburn}
u_t(\Delta) = \operatorname{max} ( u_{\text{lam}}, v(\Delta), v_{RT}),
\end{equation}
with $v(\Delta) = \sqrt {2q}$ and $v_{RT} \propto \sqrt {g \Delta}$,
where $g$ is the local gravitational acceleration.

\section{Application to the supernova problem}
 
We have carried out numerical simulations in cylindrical rather 
than in spherical coordinates, mainly because it is much simpler to
implement the level set on a Cartesian (r,z) grid. Moreover, the 
CFL condition is somewhat relaxed in comparison to spherical coordinates.
The grid we used in most of our simulations 
maps the white dwarf onto 256$\times$256 mesh points,
equally spaced for the innermost 226$\times$226 zones by
$\Delta$=1.5$\cdot$10$^6$cm, but increasing by 10\% from zone to zone in
the outer parts. The white dwarf, constructed in hydrostatic
equilibrium for a realistic equation of state, has a central density
of 2.9$\cdot$10$^9$g/cm$^3$, a radius of 1.5$\cdot$10$^8$cm, and a
mass of 2.8$\cdot$10$^{33}$g, identical to the one used by
\cite{NH95a}. The initial mass fractions of C and O
are chosen to be equal, and the total binding energy turns out to be
5.4$\cdot$10$^{50}$erg. At low densities ($\rho \leq 10^7$g/cm$^3$), 
the burning velocity of the front is set
equal to zero because the flame enters the distributed regime and our
physical model is no longer valid. However,
since in reality some matter may still burn 
the energy release obtained in the simulations is
probably somewhat too low.
 
\begin{figure*}
\begin{tabular}{cc}
{\epsfig{file=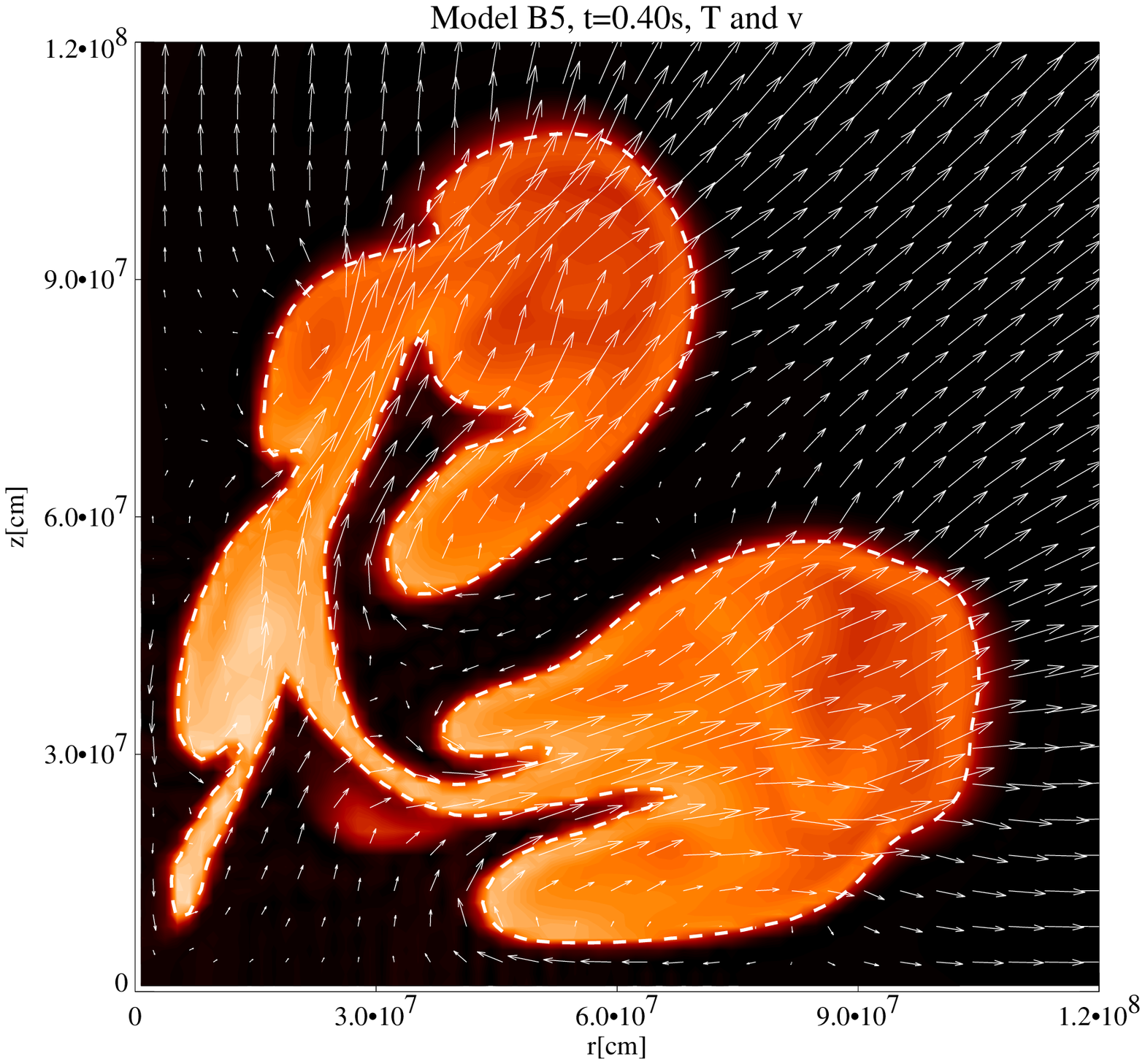, height=0.25\textheight}}
&
{\epsfig{file=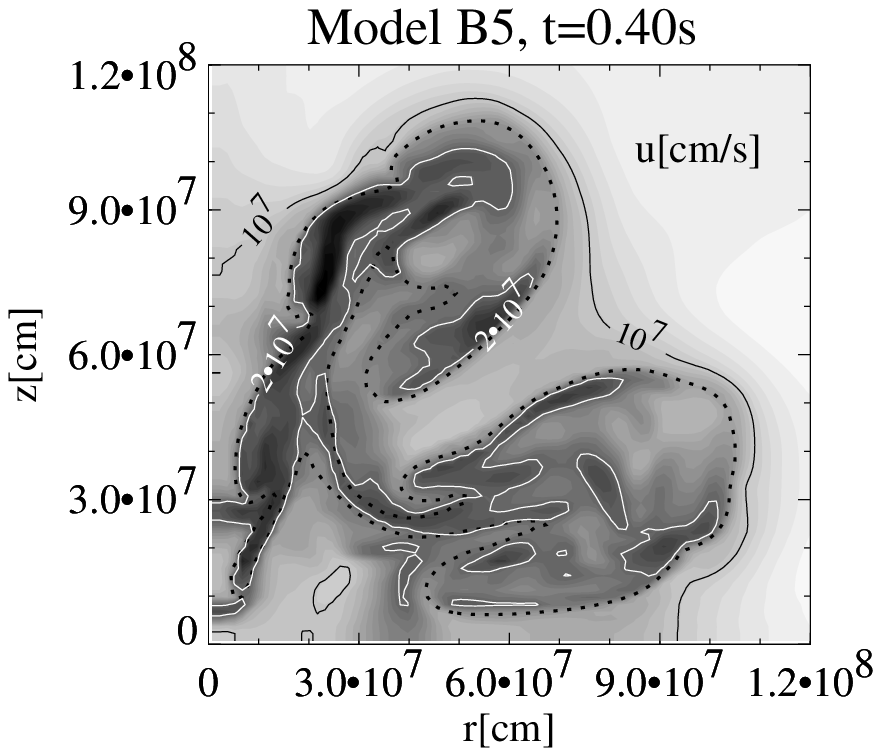, height=0.25\textheight}}
\end{tabular}
\caption{Snapshots of the temperature and the front geometry at 0.4s
for model B5 of Reinecke et al. (1999a) (right figure) and turbulent
velocity fluctuations on the grid scale (left panel). The position of
of the front is indicated by the dotted curve.}
\label{turb}
\end{figure*}

\begin{figure*}
\begin{tabular}{cc}
{\epsfig{file=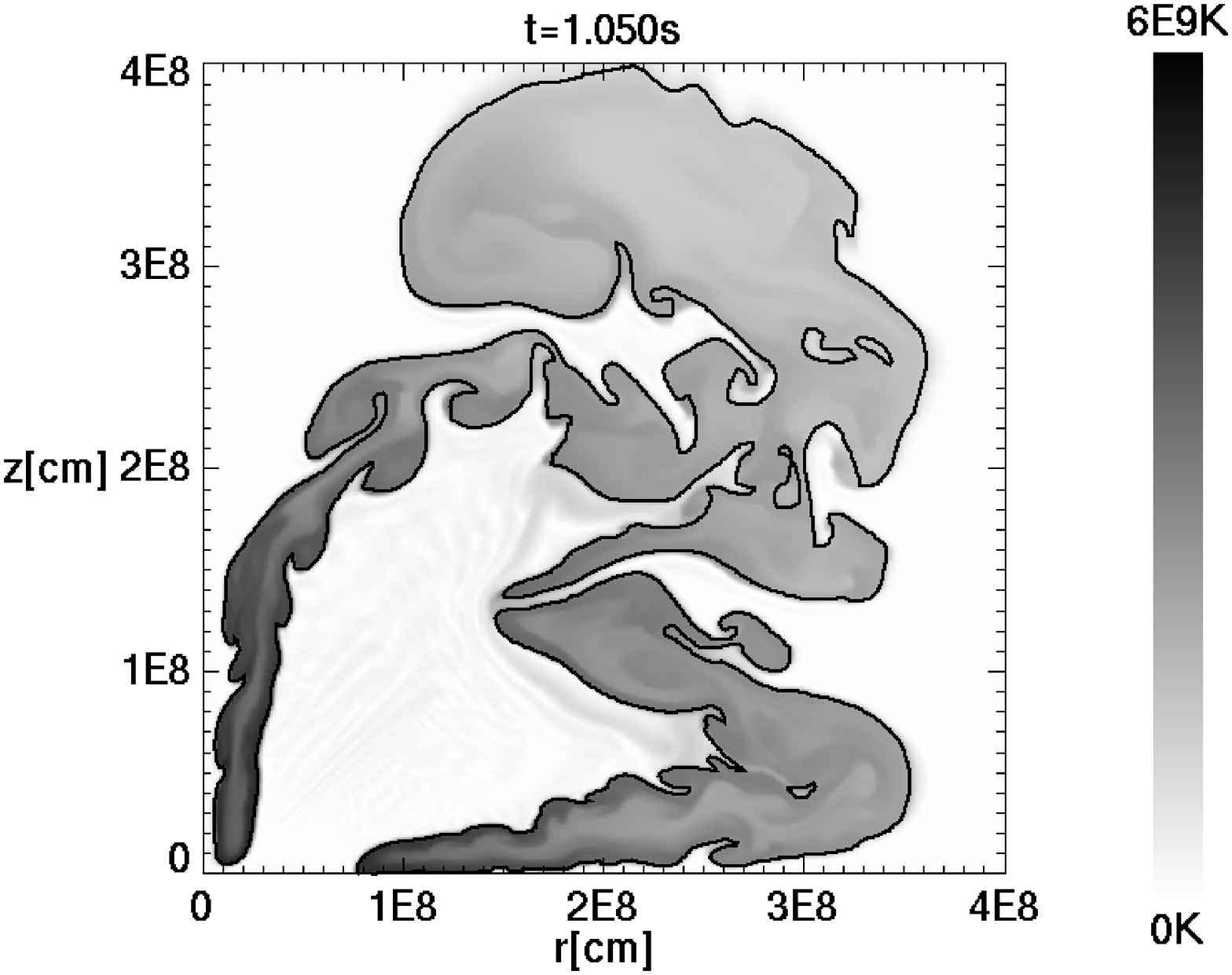, height=0.22\textheight}}
&
{\epsfig{file=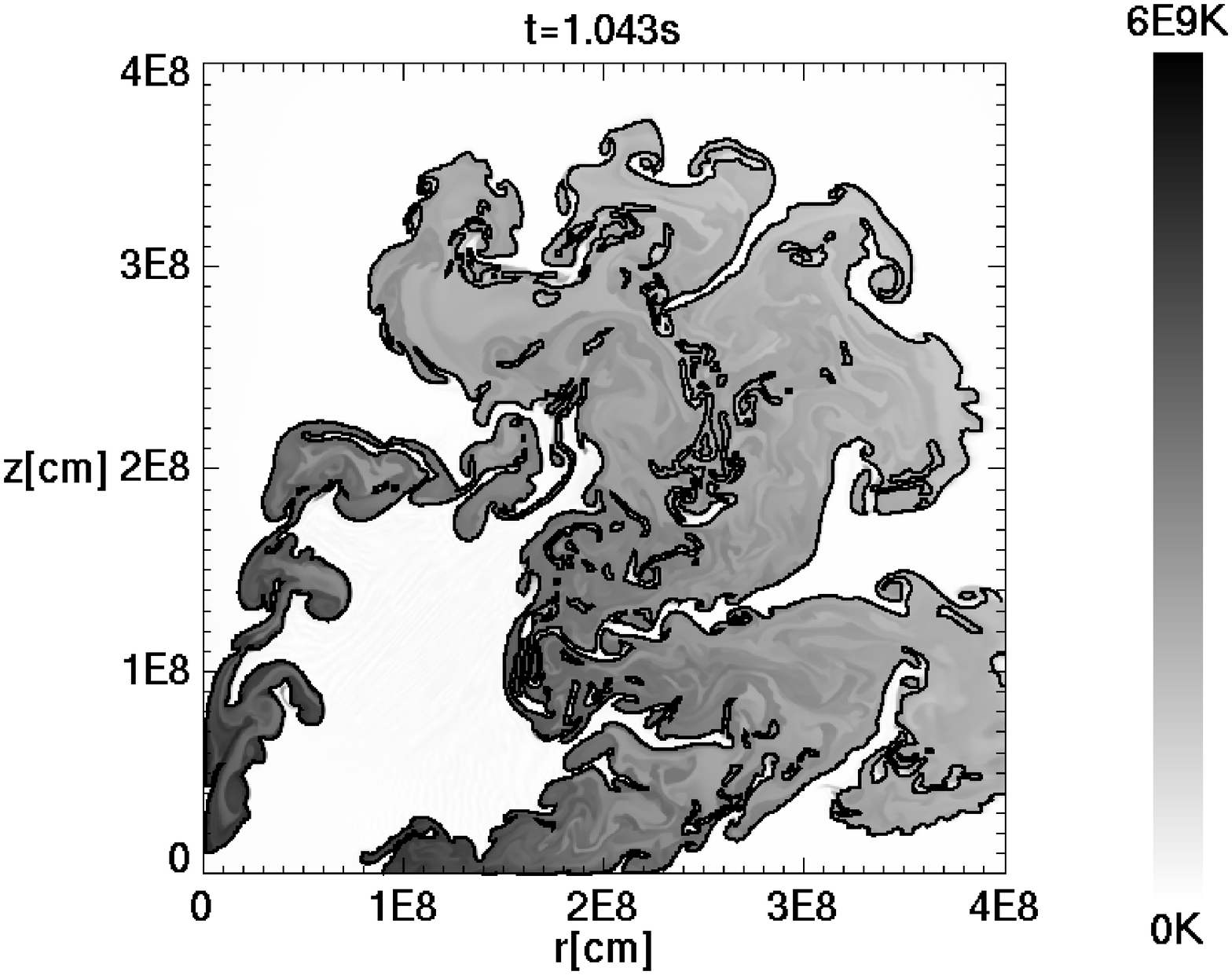, height=0.22\textheight}}
\end{tabular}
\caption{Snapshots of the temperature and the front geometry at 1.05s,
taken for the low-resolution model (left figure) and high-resolution
run, respectively.}
\label{snapsh}
\end{figure*}

Here we present only the results for two models, one in which nuclear burning was
ignited off-center in a blob and a second one in which initially five blobs
were burned as an initial condition, and refer 
to \cite{RHN99} for simulations with other initial conditions.
 
First, in Fig.\ref{turb} a snapshot of the "five-blob" model B5 is shown
at t = 0.4s. The leftt panel gives the position of the burning front,
the temperatures (in gray-shading), and the expansion velocities.
The rightpanel shows the distribution of turbulent velocity fluctuations.
We find that, in accord with one's intuition, most of the turbulence is
generated in a very thin layer near the front. Since in the limit
of high turbulence intensity the nuclear flames propagate with the
turbulent velocity it is obvious that this propagation velocity
exceeds the laminar flame speed. However, for most of the initial 
conditions we have investigated this increase was not sufficient to
unbind the star. In fact, model B5 of \cite{RHN99} was the only one
of the set that did explode.

Moreover, we show the results obtained with 3 times higher resolution
for comparison. Fig.\ref{snapsh} gives a snapshot taken at 1.05s after
ignition for the low-resolution run (left figure) and the
high-resolution run, respectively. Although the increase in spatial
resolution is only a factor of 3, the right panel shows clearly more
structure. This is an important effect because in the flamelet regime 
the rate of fuel consumption, in first order, 
increases proportional to surface area of the burning front.
The net effect is that the low-resolution model stays bound at the
end of the computations, whereas the better resolved model explodes
with an explosion energy of about 2$\times$10$^{50}$erg. 
Fig.\ref{snapsh} also demonstrates  that the level-set prescription
allows to resolve the structure of the burning front down almost
to the grid scale, thus avoiding artificial smearing of the front
which is an inherent problem of front-capturing schemes. We want to
stress that this gain of accuracy is not obtained at the expense of smaller
CFL time steps because in our hybrid scheme the hydrodynamics is still
done with cell-averaged quantities.

\section{Conclusions}
 
In this article, we have discussed the physics of thermonuclear
combustion in degenerate dense matter of C+O white dwarfs.
It was argued that not all relevant length-scales of this problem
can be numerically resolved and, therefore,
we have presented a numerical model to describe deflagration fronts
with a reaction zone much thinner then the cells of the computational 
grid. The new approach was applied to the simulate thermonuclear
supernova explosions. Our code works in 2 and 3 dimensions, but has only
been used so far in 2D.  

An implicit assumption of our numerical model is that on the resolved scales
the flows are turbulent allowing us to describe the physics on the
unresolved scales by a sub-grid model, in the spirit of large-eddy
simulations. The results presented here indicate 
that for supernova simulations this
assumption is still not satisfied if (in 2D and cylindrical
coordinates) we use a 256 $\times$ 256 grid. In fact, we could
demonstrate that for a particular set of initial conditions
increasing the numerical resolution by a factor of 3 only changes a
model that remains bound after most of the nuclear fuel was burnt into
a, although weak, explosion. The reason is simply that more structure 
on small length scales increases the rate of fuel consumption and
we suspect that even our better resolved simulation does not
yet give the final answer, pushing even 2D supernova simulation
to the limits of what can be done on present supercomputers.  

The numerical scheme outlined in this paper can, of course, also be
applied to chemical combustion in the flamelet regime and first
successful attempts to model hydrogen combustion in air have already 
been made \cite{SMK97,RHNe99}.
Its advantage is always that it resolves the flame-front down to the
grid-scale without making the time-steps intolerably small.
   
\begin{acknowledgment}
This work was supported in part by the Deutsche Forschungsgemeinschaft under
Grant Hi 534/3-1, the DAAD, and by DOE under contract No. B341495 at the University of
Chicago. The computations were performed at the Rechenzentrum Garching
on a Cray J90.
\end{acknowledgment}

 

\end{article}
\end{document}